\documentclass[aps,prc,twocolumn,superscriptaddress]{revtex4-1}

\usepackage{graphicx}
\usepackage{amsmath}
\usepackage{braket}
\usepackage{url}

\usepackage[colorlinks,
linkcolor=blue,
anchorcolor=red,
urlcolor=red,
citecolor=red]{hyperref}

\begin{document}


\title{Calculation of the Thomas-Ehrman shift in $^{16}$F and $^{15}$O(p,p) cross section with the Gamow shell model}



\author{N. Michel}\email[]{nicolas.michel@impcas.ac.cn}
\author{J. G. Li}\email[]{jianguo\_li@impcas.ac.cn}
\author{L. H. Ru}\email[]{rulonghui@impcas.ac.cn}
\author{W. Zuo}\email[]{zuowei@impcas.ac.cn}

\affiliation{CAS Key Laboratory of High Precision Nuclear Spectroscopy, Institute of Modern Physics,
Chinese Academy of Sciences, Lanzhou 730000, China}
\affiliation{School of Nuclear Science and Technology, University of Chinese Academy of Sciences, Beijing 100049, China}



\date{\today}

\begin{abstract}
The $^{16}$F nucleus is situated at the proton drip-line and is unbound by proton emission by only about 500 keV. Continuum coupling is then prominent in this nucleus. Added to that, its low-lying spectrum consists of narrow proton resonances as well. It is then a very good candidate to study nuclear structure and reactions at proton drip-line. The low-lying spectrum and scattering proton-proton cross section of $^{16}$F have then been calculated with the coupled-channel Gamow shell model framework for that matter using an effective Hamiltonian. Experimental data are very well reproduced, as well as in its mirror nucleus $^{16}$N. Isospin-symmetry breaking generated by the Coulomb interaction and continuum coupling explicitly appears in our calculations. In particular, the different continuum couplings in $^{16}$F and $^{16}$N involving $s_{1/2}$ partial waves allow to explain the different ordering of low-lying states in their spectrum. 
\end{abstract}


\maketitle

\section{Introduction}  
The area of the nuclear chart for which $A \sim 20$ possesses interesting features by many aspects. On the one hand, it can be reached from proton to neutron drip-lines with current facilities, so that comparison with experimental data can be conveniently effected \cite{RevModPhys.84.567,Motobayashi2014,TANIHATA2013215,PhysRevLett.109.202503}. On the other hand, nuclei in this region can be theoretically studied with shell model frameworks, and in particular with shell model including continuum coupling \cite{PhysRevLett.94.052501,BENNACEUR200075,0954-3899-36-1-013101}. Thus, nuclear structure at drip-lines can be precisely assessed, because the presence of a complex nuclear structure in weakly bound and unbound nuclei demands the inclusion of continuum coupling.

$^{16}$F is a light nucleus at proton drip-line and has been studied experimentally in several situations. $^{16}$F could be generated in transfer reactions involving light nuclei (see Ref.~\cite{PhysRevC.89.054315} and references therein) and the cross section of the $^{15}$O(p,p) reaction has been measured as well \cite{PhysRevC.90.014307}. $^{16}$F, as well as its mirroring nucleus $^{16}$N, are also important from an astrophysical point of view, as nearby nuclei such as $^{15}$N, $^{15-18}$O and $^{17-19}$F enter CNO cycles \cite{CNO_cycles}. 

However, due to the fact that it is unbound at ground state level, $^{16}$F is difficult to study theoretically with standard approaches, such as harmonic oscillator shell model. In fact, the model of choice for that matter is the Gamow shell model (GSM), because GSM can treat unbound nuclei bearing a complex nuclear structure \cite{PhysRevLett.89.042502,PhysRevC.67.054311,PhysRevLett.97.110603,PhysRevLett.127.262502,PhysRevC.70.064313,PhysRevC.84.051304,0954-3899-36-1-013101,PhysRevC.88.044318,PhysRevC.96.054316,PhysRevC.96.024308,PhysRevC.96.054322,PhysRevC.102.024309,PhysRevLett.121.262502,PhysRevC.103.034305,PhysRevC.104.L061306,PhysRevC.104.L061301,PhysRevC.104.024319,PhysRevC.103.L031302,Michel_Springer}. Added to that, the coupled-channel GSM (GSM-CC) allows to calculate reaction cross sections in the GSM framework using the same Hamiltonian, so that both nuclear structure and reaction observables can be assessed at the same time \cite{PhysRevC.89.034624,PhysRevC.99.044606,JPG_GX_Dong,PhysRevC.91.034609,Michel_Springer}. Isospin-symmetry breaking can then be quantitatively considered, as has been done, for example, in oxygen and carbon isotopes and isotones in Refs.~\cite{PhysRevC.100.064303,PhysRevC.103.044319}. We will then use GSM and GSM-CC to calculate the spectrum of $^{16}$F, that of its mirror nucleus $^{16}$N, and the excitation function of the $^{15}$O(p,p) reaction.

This paper is then structured as follows. We will firstly present the basic features of GSM and GSM-CC. Then, we will depict the model spaces and Hamiltonian used for the theoretical description of the spectra of $^{16}$F and $^{16}$N and the calculation of the cross section of the $^{15}$O(p,p) reaction. We will then show obtained results and comment on the isospin asymmetry obtained in the spectra of $^{16}$F and $^{16}$N, in particular. Conclusion will be made afterwards.

\section{GSM and GSM-CC models}

The fundamental equation in GSM is the one-body Berggren completeness relation of a given partial wave of quantum numbers $\ell,j$ \cite{BERGGREN1968265}. It reads:
\begin{equation}
\sum_n{u_n(r) u_n(r')} + \int_{L^+} u_k(r) u_k(r')~dk = \delta(r-r') \label{Berggren}
\end{equation}
where $u_n(r)$ is a bound or resonance state and $u_k(r)$ is a scattering state belonging to the complex contour of one-body momenta $L^+$. The $L^+$ contour must encompass all the resonances present in the discrete sum of Eq.(\ref{Berggren}) (see Refs.\cite{0954-3899-36-1-013101,Michel_Springer} for details).

In order to be able to use Eq.(\ref{Berggren}) in numerical calculations, its integration contour must be discretized. For this, one uses the Gauss-Legendre quadrature, which has been seen to converge quickly as about 15 points per contour are sufficient to obtain convergence for energies \cite{PhysRevC.83.034325}. Then, one obtains the discretized Berggren completeness relation, which is then formally identical to the harmonic oscillator completeness relation. Consequently, one can build Slater determinants from the discretized Berggren completeness relation to obtain a basis of many-body states similarly to harmonic oscillator shell model. The main difference is that the GSM Hamiltonian matrix becomes complex symmetric, where numerous many-body scattering states are present. Special numerical techniques then had to be devised to diagonalize the GSM Hamiltonian matrix, which are based on the use of the Jacobi-Davidson and the overlap methods \cite{Jacobi_Davidson,0954-3899-36-1-013101,MICHEL2020106978,Michel_Springer}. As these techniques have been thoroughly explained in these citations, we refer the reader to these papers and book for details.

While GSM allows to calculate many-body halo and resonances, it is not sufficient if one aims at evaluating reaction cross sections. This arises because the many-body scattering eigenstates of the GSM matrix do not have a well defined asymptotic behavior, i.e.~they consist of a complex linear combination of reaction channels of a given energy. As it is necessary to deal with well defined reaction channels to be able to calculate reaction observables, GSM-CC has been developed for that matter \cite{PhysRevC.89.034624,PhysRevC.99.044606,JPG_GX_Dong,PhysRevC.91.034609,Michel_Springer}. 

In GSM-CC, the many-body resonant or scattering coupled-channel states $\ket{ \Psi_{ M_{\rm A} }^{ J_{\rm A} } }$ to determine are built from targets and projectiles issued from a GSM calculation :
 \begin{equation}
      \ket{ \Psi_{ M_{\rm A} }^{ J_{\rm A} } } = \sum_{c} \int_{0}^{+\infty} \left( \frac{u_c(r)}{r} \right)   \ket{ (c,r) } { r }^{ 2 } dr \ ,	
      \label{scat_A_body_one_nuc}
    \end{equation}
    where $\ket{ \Psi_{ M_{\rm A} }^{ J_{\rm A} }}$ is the resonant or scattering state of $A$ nucleons, $u_c(r)$ is the radial channel wave function to determine and $\ket{ (c,r) }$ is the reaction channel :
 \begin{equation}
 \ket{ (c,r) } = \hat{\mathcal{A}} { \{ \ket{ \Psi_{\rm T}^{ J_{\rm T} } } \otimes \ket{ r \ell j } \} }_{ M_{\rm A} }^{ J_{\rm A} } \label{c_channel}
    \end{equation}
    where $\ket{r \ell j}$ is a nucleon projectile state of fixed quantum numbers $\ell,j$ and $\ket{ \Psi_{\rm T}^{ J_{\rm T}}}$ is the $A-1$ target state, eigenstate of the GSM Hamiltonian, both being coupled to $J_A,M_A$, so that $c$ embodies all the quantum numbers of the considered $c$ channel.

   The coupled-channel equations of GSM-CC are derived from the Schr\"odinger equation $H \ket{ \Psi_{ M_{\rm A} }^{ J_{\rm A} } } = E \ket{ \Psi_{ M_{\rm A} }^{ J_{\rm A} } }$, with $E$ the energy of the many-body coupled-channel state :
   \begin{equation}
      \sum_{c} \int_{0}^{+\infty} \left( H_{ cc' } (r,r') - E ~ N_{ cc' } (r,r') \right) u_c (r') dr' = 0	\ ,
      \label{cc_eq}
    \end{equation}
    with
    \begin{align}
      & H_{ cc' } (r,r') = rr'\bra{ (c,r) } \hat{H} \ket{ (c',r') } \label{h_ccp} \\
      & N_{ cc' } (r,r') = rr'\braket{ (c,r) | (c',r') }.  
      \label{n_ccp}
    \end{align}

The matrix elements of the GSM-CC coupled-channel equation in Eqs.(\ref{h_ccp},\ref{n_ccp}) are conveniently calculated by expanding $\ket{r \ell j}$ in the Berggren basis in Eq.(\ref{c_channel}):
 \begin{equation}
    \hat{\mathcal{A}} [\ket{\Psi_{\rm T}} \otimes \ket{r \ell j}]^{J_{\rm A}}_{M_{\rm A}}  = \sum_n \frac{u_n(r)}{r} \{a^{\dag}_{n \ell j} \ket{ \Psi_{\rm T}^{ J_{\rm T} } } \}^{J_{\rm A}}_{M_{\rm A}}
    \label{Berggren_basis_channels}
\end{equation}
Indeed, as $\ket{\Psi_{\rm T}}$ is a linear combination of Slater determinants expressed in the same Berggren basis as projectiles, antisymmetry is straightforward to impose in Eq.(\ref{Berggren_basis_channels}) with creation and annihilation operators. GSM-CC matrix elements are then obtained via Slater determinant algebra \cite{Michel_Springer}. 

As $c$ channels are non-orthogonal, it is more convenient to deal with orthogonalized channels.
Orthogonalized channels are obtained by diagonalizing the overlap matrix $N_{ cc' } (r,r')$
of Eq.(\ref{n_ccp}) in the Berggren basis \cite{Michel_Springer}. Let us then denote by $w_c(r)$ the channel radial wave functions associated to orthogonalized channels. The channels functions $w_c(r)$ then proceed from coupled-channel equations, whose structure is the same as those occurring in standard reaction theory:
      \begin{eqnarray}
        &&\left( \frac{ \hbar^2 }{ 2 \mu } \left( -\frac{d^2}{dr^2} + \frac{ \ell_c ( \ell_c + 1 ) }{ {r}^{2} } \right) + {V}_{c}^{ (\rm loc) } (r) \right) {w}_{c} (r) \nonumber \\
        &+& \sum_{ c' } \int_{0}^{ +\infty } V_{ c c' }^{ (\rm non-loc) } (r , r') ~ w_{c'}(r') ~ dr' \nonumber \\
        &&= ( E - {E}_{\rm T_{c} }) w_{c} (r)    \ ,
        \label{cc_eqs_modif_Eqs_full_2}
      \end{eqnarray}
where $\mu$ is the effective mass of the projectile, $\ell_c$ is the channel orbital momentum, $E_{{\rm T}_c}$ is the channel target energy and ${V}_{c}^{ (\rm loc) } (r)$, $V_{ c c' }^{ (\rm non-loc) } (r , r') $ are the local and non-local coupled-channel potentials issued from Eqs.(\ref{h_ccp},\ref{n_ccp}), respectively. 

Eq.(\ref{cc_eqs_modif_Eqs_full_2}) is an integro-differential coupled-channel equation and is strongly non-local. It cannot be integrated in coordinate space with iterative methods, such as the equivalent potential method \cite{VAUTHERIN1967175,Michel09}, due to numerical instability. Thus, we developed a technique of resolution based on Berggren basis expansion \cite{PhysRevC.99.044606,Michel_Springer}. Equation (\ref{cc_eqs_modif_Eqs_full_2}) then reduces to a matrix diagonalization problem for bound and resonance states and to a linear system to solve for scattering states, so that Berggren basis expansion is hereby very stable.
Initial $u_c(r)$ functions are then obtained from $w_c(r)$ functions via the matrix transformation induced by $N_{ cc' } (r,r')$. For a more thorough presentation of GSM-CC, we refer the reader to Ref.\cite{Michel_Springer}.

\section{Hamiltonian and model space}
The used framework is that of the core + valence nucleons picture, where the core consists of the $^{12}$C nucleus. The $^{12}$C core is mimicked by a Woods-Saxon (WS) potential. Valence nucleons interact via the Furutani-Horiuchi-Tamagaki (FHT) nuclear interaction \cite{FHT1,FHT2,PhysRevC.96.054316}, to which the Coulomb interaction is added for protons. The model space is constructed from $spd$ partial waves.
Note that this model has been successfully used for the description of the reaction $^{14}$O(p,p) reaction, which was the object of a recent experiment \cite{PhysRevLett.121.262502}.

As the neutron part is well bound in $^{16}$F, it only consists of the harmonic oscillator shells $0p_{1/2}$, $0d_{5/2}$, $1s_{1/2}$ and $0d_{3/2}$, whose oscillator length is $b$ = 2 fm. The proton part is represented by the Berggren basis, so that it consists of the $S$-matrix poles $0p_{1/2}$, $0d_{5/2}$, $1s_{1/2}$ and $0d_{3/2}$, to which complex-momentum scattering contours of orbital angular momentum $\ell \leq 2$ are added. The latter all start at $k=0$, are peaked in $k=0.2-i0.1$ fm$^{-1}$, return to the real $k$-axis in $k=0.4$ fm$^{-1}$ and end in $k=2$ fm$^{-1}$. They are discretized with 15 points each. Not more than three nucleons are allowed to occupy non-resonant shells.

As we consider the core + valence nucleons approach, the most natural framework to define particle coordinates is the cluster orbital shell model (COSM) \cite{Suzuki1988}. All valence nucleons coordinates are defined with respect to the center of mass of the core, so that they are translationally invariant and no center of mass excitation can occur. The recoil of the core is taken into account in the Hamiltonian by an additional two-body kinetic term (see Ref.\cite{Michel_Springer} for more details about the COSM framework applied to GSM). The COSM Hamiltonian then reads:
\begin{equation}
H = \sum_i \left( \frac{\mathbf{\hat{p}}^2_i}{2 \mu_i} + \hat{U}_i \right) + \sum_{i<j \in val} \left( \hat{V}_{ij} + \frac{\mathbf{\hat{p}}_i \cdot \mathbf{\hat{p}}_j}{M_{core}} \right) \label{H}
\end{equation}
where $\mu_i$ is the effective mass of the nucleon $i$, $\hat{U}_i$ is the core potential acting on the nucleon $i$, $\hat{V}_{ij}$ is the FHT+Coulomb interaction and the last term is the recoil term, inversely proportional to the mass of the core $M_{core}$.

The parameters of the Hamiltonian of Eq.(\ref{H}) have been fitted from the low-lying spectra of $^{14}$N (0$^+$ and 1$^+$ states), $^{15}$O (1/2$^-$, 1/2$^+$ and 5/2$^+$ states), and $^{16}$F (0$^-$-3$^-$ and 1$^+$-3$^+$ states).
For this, we firstly fitted the WS core potential depths and FHT interaction parameters on these states by having two nucleons at most in scattering shells.
This resulted in residues of typically 500 keV to 1 MeV per considered state. Afterwards, we refitted the WS core potential depths only by letting three particles occupy non-resonant basis states, thus focusing only on $^{16}$F separation energy and spectrum.
This doing, their typical residues could reach 100-200 keV.

Let us now enumerate the parameters fitted with the above fitting procedure. The WS core potential has a diffuseness $d$=0.65 fm, a radius $R_0$ = 2.907 fm, and potential depths depend on both nucleon type and orbital angular momentum :
for protons, one has $V_o = $ 34.716 MeV, 22.598 MeV, and 37.032 MeV, respectively for $\ell=0,1,2$, whereas for neutrons,
one has $V_o = $ 56.699 MeV, 57.851 MeV, and 61.815 MeV, also respectively for $\ell=0,1,2$. Spin-orbit potential depths read : for protons, one has $V_{so} = $ 9.514 MeV, and 4.8 MeV,
respectively for $\ell=1,2$, while for neutrons, one has $V_{so} = $ 8.885 MeV and 7.129 MeV, respectively for $\ell=1,2$.
The parameters of the FHT interaction are listed in Tab.(\ref{FHT}).

The spin-orbit component of the FHT interaction is equal to zero (see Tab.(\ref{FHT})).
We suppressed it because its overall effect was seen to be redundant in the Hamiltonian. Indeed, along with the spin-orbit part of the WS core potential, spin-orbit degrees of freedom are present via the tensor part of the FHT interaction.
Hence, we could put the spin-orbit component of the FHT interaction to zero while still retaining almost all spin-orbit degrees of freedom in the Hamiltonian.

\begin{table}[htb] 
\centering
\caption{\label{Table.InterParamFHT} The optimized parameters of the FHT interaction consist of central ($V_c^{ST}$) and tensor ($V_T^{ST}$) coupling constants \cite{PhysRevC.96.054316}.
$S=0,1$ and $T=0,1$ are the spin and isospin of the two nucleons, respectively.
Parameters are given in MeV for the central part and in MeV fm$^{-2}$ for the tensor part.
The spin-orbit  component of the FHT interaction has been arbitrarily put to zero. \label{FHT} }
\begin{tabular}{l|cccccc}\hline \hline 
Parameter &$V_c^{11}$  & $V_c^{10}$ &$V_c^{00}$ & $V_c^{01}$ & $V_{T}^{0}$  & $V_{T}^{1}$  \\ \hline 
Value&  12.371 &  $-$15.934  & $-$0.362 &  $-$15.248 & $-$0.091 & $-$21.207 \\ \hline \hline
\end{tabular}
\end{table}

From a theoretical point of view, i.e.~without model space truncation and including all reaction channels, the Hamiltonian used in GSM and GSM-CC is the same.
Thus, eigenstates should have identical energies when calculated either with GSM or GSM-CC.
Nevertheless, this is not the case in practice due to the different truncation schemes used in both model spaces. In order to compensate for this difference, and also for GSM-CC results to reproduce experimental data as close as possible, the two-body matrix elements of the nuclear interaction are multiplied by a factor close to one in GSM-CC. They are equal to 1.006, 1.009, 1.003 and 1.004 for the channels of quantum numbers $J^\pi=0^-$, $1^-$, $2^-$ and $3^-$, respectively. The induced change is minimal and obviously does not modify the initial physical properties of the Hamiltonian fitted with GSM. No corrective factor is introduced for other channels. 

As $^{16}$F and $^{16}$N are mirror nuclei, the model space and Hamiltonian used for $^{16}$N are obtained from those of $^{16}$F by exchanging protons and neutrons.
Thus, the neutron part of the model space of $^{16}$N consists of partial waves represented by the Berggren basis, whereas its proton part consists of harmonic oscillator shells.
The parameters of the WS core neutron potential of $^{16}$N are then those of the WS core proton potential of $^{16}$F (same for the WS core proton potential of $^{16}$N).
This Hamiltonian definition allows to assess the Thomas-Ehrman shift because $^{16}$F and $^{16}$N would have the same wave functions and energies up to proton-neutron exchange in the absence of Coulomb interaction (see also Refs.\cite{PhysRevC.100.064303,PhysRevC.103.044319} for studies of isospin-symmetry breaking with GSM).

\section{Calculations and discussions} 
\begin{figure}
    \centering
    \includegraphics[width=1.0\columnwidth]{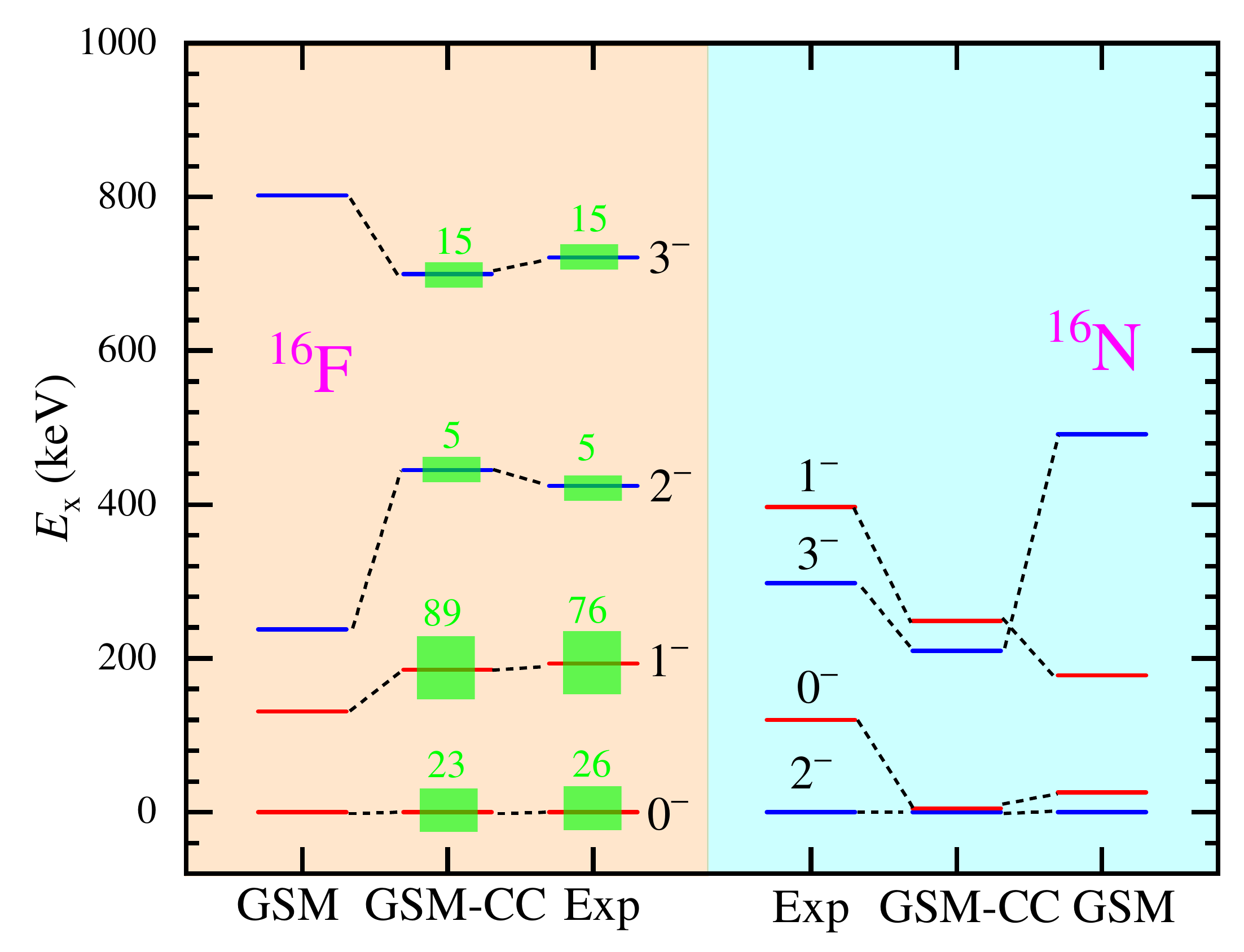}
    \caption{Low-lying energy spectra of the $^{16}$F and $^{16}$N nuclei obtained with GSM and GSM-CC and compared to experimental data (Exp). The proton-emission widths are indicated by filled squares and their numerical value is also indicated above. Excitation energies $E_x$ and proton-emission widths are given in keV. Experimental data are taken from Ref.\cite{ensdf}.}
    \label{fig:spectra}
\end{figure}
The spectra of $^{16}$F and $^{16}$N are shown in Fig.~\ref{fig:spectra}. A good description of eigenenergies is obtained with GSM in both systems, as the maximal deviation from experiment is around 200 keV. While the ordering of $^{16}$F excitation energies is reproduced with GSM, the $1^-$ and $3^-$ energies are inverted in $^{16}$N. However, one can see that the Coulomb interaction is responsible for the different orderings of the $0^-$, $1^-$ and $2^-$ states in $^{16}$F and $^{16}$N. Indeed, the ground state and first excited states of $^{16}$N are found to be $2^-$ and $0^-$ states, respectively, as is the case experimentally. Nevertheless, the proton-emission widths of $^{16}$F eigenstates are almost zero with GSM. The reason for this probably comes from model space truncation and contour discretization effects, as a very large model space would be necessary to obtain a precision of 1-5 keV for complex energies.

As excitation energies are shown in Fig.~\ref{fig:spectra}, ground state energies are put to zero, so that one cannot assess the error made on the $^{16}$N ground state binding energy. In fact, it is about 830 keV too bound with respect to experimental data, which results from the proton-neutron interchange defining its Hamiltonian. But, as $^{16}$N is well bound, this does not change any of its physical properties.

Corrective factors have been introduced at GSM-CC level to reproduce experimental data in $^{16}$F, so that the energy spectrum of $^{16}$F is almost exactly reproduced with GSM-CC (see Fig. \ref{fig:spectra}). Interestingly, the ordering of excited states of $^{16}$N is also recovered at GSM-CC level, even though the $2^-$ and $0^-$ states are close to be degenerate (see Fig. \ref{fig:spectra}). This was rather unexpected as the same corrective factors are used in $^{16}$F and $^{16}$N. GSM-CC excitation energies in $^{16}$N are also close to experimental data, as the typical error is about 150 keV. The proton emission widths of $^{16}$F eigenstates are very well reproduced, as only the width of the $1^-$ state is larger by a little more than 10 keV compared to experiment, the others differing by at most 3 keV. This shows that the wave functions of $^{16}$F eigenstates consist principally of channels made of well bound $^{15}$O states + an unbound proton.

The most striking consequence of isospin-symmetry breaking in weakly bound and resonance nuclear states is the Thomas-Ehrman shift, i.e.~the different excitation energies of the nuclei belonging to the same isospin multiplet in the absence of Coulomb interaction \cite{PhysRev.88.1109,PhysRev.81.412}. The Thomas-Ehrman shift is clearly visible in the low-lying spectrum of $^{16}$F and $^{16}$N (see Fig.~\ref{fig:spectra}). It is generated by two factors: the presence of the additional Coulomb interaction in $^{16}$F and the unbound character of $^{16}$F, whereby $^{16}$N is well bound. The Thomas-Ehrman shift is especially well reproduced in GSM-CC, and the effect of the Coulomb interaction + continuum coupling is also present in GSM. 

In order to further determine the origin of the Thomas-Ehrman shift in $^{16}$F and $^{16}$N, we calculated the average values of the different kinetic, nuclear and Coulomb parts entering the GSM Hamiltonian. We could see that the expectation values of the kinetic operator and nuclear interaction for protons in $^{16}$F is about 8 MeV smaller in the $0^-$ and $1^-$ states than in the $2^-$ and $3^-$ states. Conversely, this value is close to 5 MeV in $^{16}$N for its neutron part. The expectation values of the other operators entering the Hamiltonian are close for all $^{16}$F states (same for $^{16}$N states). 

This indicates that the Thomas-Ehrman shift is mainly due to the Hamiltonian matrix elements pertaining to $s_{1/2}$ partial waves in GSM and GSM-CC. Indeed, as the $1s_{1/2}$ proton state in $^{16}$F is above particle-emission threshold, both its average kinetic energy and radial wave function inside the nucleus are smaller in magnitude than those of the well bound $1s_{1/2}$ neutron state in $^{16}$N. Consequently, kinetic and nuclear matrix elements associated to the $1s_{1/2}$ proton state in $^{16}$F can be expected to be smaller in magnitude than those related to the $1s_{1/2}$ neutron state in $^{16}$N.
Added to that, one observed a larger continuum coupling in the $0^-$ and $1^-$ states in $^{16}$F than in its $2^-$ and $3^-$ states, as it is about 0.35 in the former and 0.2 in the latter.
This accentuates the differences between configurations containing $s_{1/2}$ partial waves involving either protons in $^{16}$F or neutrons in $^{16}$N.
Consequently, the $0^-$ and $1^-$ states are close in $^{16}$F because of their similar structure, whereby the average proton occupation of the $s_{1/2}$ partial wave is close to one. Conversely, it is about 0.1 in the $2^-$ and $3^-$ states of $^{16}$F. This also explains why the $0^-$ and $1^-$ states are below the $2^-$ and $3^-$ states. Indeed, as continuum coupling is stronger in the $0^-$ and $1^-$ states, they can gain more binding energy compared to the $2^-$ and $3^-$ states. This is directly seen in the GSM energy differences of these states in $^{16}$F and $^{16}$N, as it is about 6 MeV for the $0^-$ and $1^-$ states and 6.3 MeV for the $2^-$ and $3^-$ states.

It is also interesting to consider the expectation value of the $\hat{T}^2$ operator in $^{16}$F and $^{16}$N. However, the present definition of $^{16}$F and $^{16}$N model spaces is not suitable to evaluate isospin-symmetry breaking with $\hat{T}^2$, contrary to Thomas-Ehrman shift. Indeed, as proton-neutron spaces are asymmetric in $^{16}$F and $^{16}$N, a spurious isospin mixture would  occur, which would blur the physical deviation of $\hat{T}^2$ expectation values from integers. Hence, we fitted a GSM Hamiltonian in a symmetric proton-neutron space for $^{16}$F and $^{16}$N in order to have a proper estimate of isospin expectation values. We obtained that $T \simeq 1$ for all eigenstates of $^{16}$F and $^{16}$N, where deviations are of the order of $10^{-3}$ or less. Isospin-symmetry breaking is thus mainly of dynamical character \cite{LEVIATAN201193,PhysRevLett.89.222501}, i.e.~the Hamiltonians and spectra of $^{16}$F and $^{16}$N differ because of the Coulomb force and different wave function asymptotes, while the isospin of many-body eigenstates is almost $T=1$.

\begin{figure}
    \centering
    \includegraphics[width=1.0\columnwidth]{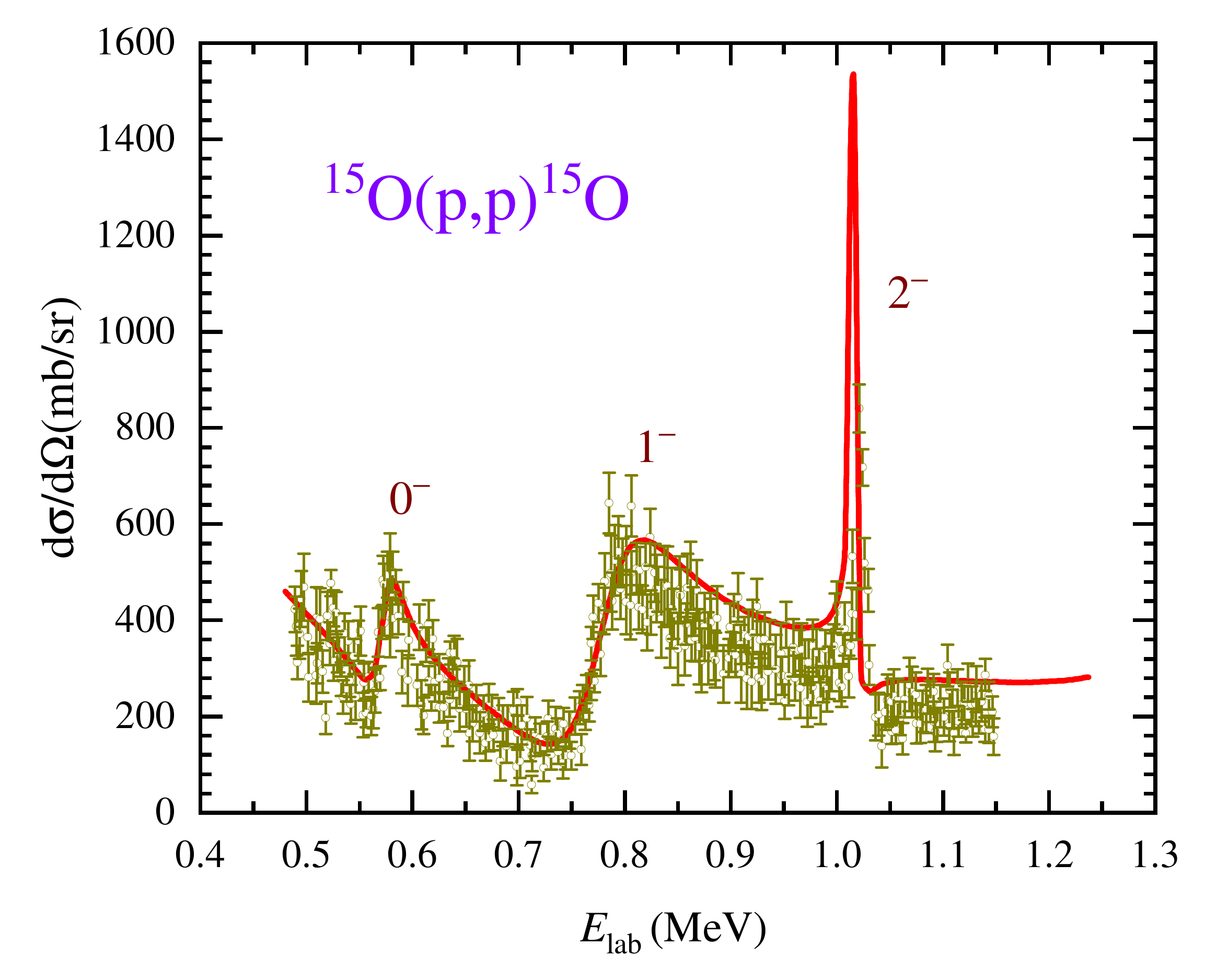}
    \caption{Excitation function of the $^{15}$O(p,p) scattering reaction calculated with GSM-CC and compared to experimental data. Energies and cross section are defined in the laboratory frame. The cross section angle is defined in the center of mass frame and is equal to 180 degrees. The low-lying resonance states of $^{16}$F are indicated on the figure next to their associated projectile excitation energy. Experimental data are taken from Ref.\cite{PhysRevC.90.014307}.}
    \label{fig:cross_section}
\end{figure}
Following the very satisfactory reproduction of the spectrum of $^{16}$F in our calculations, the excitation function of the $^{15}$O(p,p) scattering reaction has been evaluated in GSM-CC. Note that the experimental cross section was not part of the fitting process, so that its calculation is fully predictive. GSM-CC results are depicted in Fig.~\ref{fig:cross_section} along with experimental data \cite{PhysRevC.90.014307}. One can see that the experimental cross section is accurately reproduced for the whole range of projectile energies. For this, an almost exact fit of excitation energies had to be obtained in GSM-CC, and proton-emission widths had to be reproduced up to few keV compared to experimental data. Note, however, that experimental errors are fairly large, and that the GSM-CC cross section typically lies at the upper end of experimental error bars. Added to that, the GSM-CC cross section peak at the $2^-$ state is about 1.5 higher than that of the R-matrix fit of Ref.\cite{PhysRevC.90.014307}. It has been checked that an overall 0.85 renormalization factor allows for the GSM-CC cross section to be situated around the middle of experimental error bars. This probably indicates that the neglected reaction channels in the GSM-CC Hamiltonian might have an influence on the cross section and would have it slightly decreased. Nevertheless, the GSM-CC calculation clearly reproduces the experimental cross section overall and shows that the used model space recaptures all its most important physical features.

\section {Summary} 
Nuclei of the $A \sim 20$ region are interesting by many aspects. Firstly, they are accessible with current accelerator facilities from proton to neutron drip-lines. Secondly, their theoretical study can be very accurate as associated nuclear states can be treated with shell model frameworks including continuum degrees of freedom. Added to that, the Coulomb barrier is still low in this region, so that continuum effects exist at proton and neutron drip-lines, which allow to study the nucleon-nucleon interaction and isospin-breaking effects at the limit of stability. These nuclei are also of astrophysical interest, as the nuclei entering CNO cycles belong to this region of the nuclear chart.

For that matter, we studied the mirror systems $^{16}$F and $^{16}$N. While being complex systems, the structure of their low-lying states can be described in sizable model spaces. Nevertheless, as $^{16}$F is unbound by proton emission, continuum coupling must be included for its proper description, so that GSM and GSM-CC are the models of choice for that matter. 

We have thus shown that the low-lying spectrum of $^{16}$F and $^{16}$N, as well as the $^{15}$O(p,p) scattering reaction cross section, can be reproduced using GSM and GSM-CC with an effective Hamiltonian, consisting of a $^{12}$C core and valence nucleons interacting with an effective nuclear interaction. Obtained results are indeed very satisfactory, especially for $^{16}$F at GSM-CC level, where the GSM-CC cross section recaptures the most important physical features of experimental data. However, it lies at the upper limit of its large error bars, which indicates the possible influence of neglected emission channels, whose overall effect would be to slightly decrease the excitation function by the same factor at each projectile energy.

The isospin-symmetry breaking induced by both the Coulomb interaction and coupling to the continuum can be seen in the GSM and GSM-CC spectra of $^{16}$F and $^{16}$N. In particular, in our calculation, the observed Thomas-Ehrman shift occurs because of the special role played by the proton $s_{1/2}$ partial wave in $^{16}$F. On the one hand, as the $0^-$ and $1^-$ states of $^{16}$F have similar $s_{1/2}$ proton content, they are consecutive in its spectrum. On the other hand, the stronger continuum coupling in these states, being about twice as large as in the $2^-$ and $3^-$ states of $^{16}$F, provides them more binding energy. Thus, the origin of the different ordering of low-lying states in $^{16}$F compared to $^{16}$N could be identified using the GSM framework.

\vspace*{1cm}
\begin{acknowledgments}
 We than Prof.~X.~D.~Tang for useful comments.
 This work has been supported by the National Natural Science Foundation of China under Grant Nos. 12175281 and 11975282; the Strategic Priority Research Program of Chinese Academy of Sciences under Grant No. XDB34000000; the Key Research Program of the Chinese Academy of Sciences under Grant No. XDPB15; the State Key Laboratory of Nuclear Physics and Technology, Peking University under Grant No. NPT2020KFY13. This research was made possible by using the computing resources of Gansu Advanced Computing Center. 

\end{acknowledgments}

\bibliography{references.bib}
\end{document}